\documentclass[letterpaper]{article} 
\usepackage{amsfonts}
\usepackage{aaai2026}  
\usepackage{times}  
\usepackage{helvet}  
\usepackage{courier}  
\usepackage[hyphens]{url}  
\usepackage{graphicx} 
\usepackage{natbib}  
\usepackage{caption} 
\usepackage{algorithm}

\usepackage{newfloat}
\usepackage{listings}
\usepackage{amsmath,amssymb,amsfonts}
\usepackage{graphicx}
\usepackage{textcomp}
\usepackage{amsmath}
\usepackage{cases}
\usepackage{diagbox}
\usepackage{booktabs}
\usepackage{graphicx}
\usepackage{subfigure}
\usepackage{algpseudocode}
\usepackage{amsmath}
\usepackage{graphics}
\usepackage{epstopdf}
\usepackage{bm}
\usepackage{amssymb,amsfonts}
\usepackage{color}
\usepackage{float}
\usepackage{lineno}

\usepackage{url,multirow}

\usepackage{newfloat}
\usepackage{listings}
\DeclareCaptionStyle{ruled}{labelfont=normalfont,labelsep=colon,strut=off} 
\floatstyle{ruled}
\newfloat{listing}{tb}{lst}{}
\floatname{listing}{Listing}
\title{CWEFS: Brain volume conduction effects inspired channel-wise EEG feature selection for multi-dimensional emotion recognition}
\author{
    \textbf{Xueyuan Xu*}, \textbf{Wenjia Dong}, \textbf{Fulin Wei}, \textbf{Li Zhuo} \\
    {\normalfont School of Information Science and Technology, Beijing University of Technology, Beijing 100124, China} \\
    {\normalfont School of Artificial Intelligence, Anhui University, Beijing 100124, China} \\
    {\normalfont \{xxy, zhuoli\}@bjut.edu.cn, 23027425@emails.bjut.edu.cn, weifulin@ahu.edu.cn}
}

\usepackage{bibentry}

\begin{document}

\maketitle

\begin{abstract}
Due to the intracranial volume conduction effects, high-dimensional multi-channel electroencephalography (EEG) features often contain substantial redundant and irrelevant information. This issue not only hinders the extraction of discriminative emotional representations but also compromises the real-time performance. Feature selection has been established as an effective approach to address the challenges while enhancing the transparency and interpretability of emotion recognition models. However, existing EEG feature selection research overlooks the influence of latent EEG feature structures on emotional label correlations and assumes uniform importance across various channels, directly limiting the precise construction of EEG feature selection models for multi-dimensional affective computing. To address these limitations, a novel channel-wise EEG feature selection (CWEFS) method is proposed for multi-dimensional emotion recognition. Specifically, inspired by brain volume conduction effects, CWEFS integrates EEG emotional feature selection into a shared latent structure model designed to construct a consensus latent space across diverse EEG channels. To preserve the local geometric structure, this consensus space is further integrated with the latent semantic analysis of multi-dimensional emotional labels. Additionally, CWEFS incorporates adaptive channel-weight learning to automatically determine the significance of different EEG channels in the emotional feature selection task. The effectiveness of CWEFS was validated using three popular EEG datasets with multi-dimensional emotional labels. Comprehensive experimental results, compared against nineteen feature selection methods, demonstrate that the EEG feature subsets chosen by CWEFS achieve optimal emotion recognition performance across six evaluation metrics.
\end{abstract}


\section{Introduction}
Electroencephalography (EEG) is a non-invasive and portable technology that measures brain activity, enabling rapid responses to various emotional states \cite{tang2023flexible}. Recently, EEG-based emotion recognition has garnered significant attention in multimedia-induced affective computing due to its high temporal resolution and cost-effectiveness \cite{gong2023astdf,wu2023affective,li2021multi}. To characterize the non-stationary and nonlinear properties of EEG signals, numerous feature extraction methods have been developed (e.g., differential entropy (DE)\cite{Duan2013DE}, differential asymmetry (DASM)\cite{liu2013real}), facilitating more accurate representations of diverse emotional states.

\begin{figure}[!t]
\centering
\includegraphics[width=0.45\textwidth]{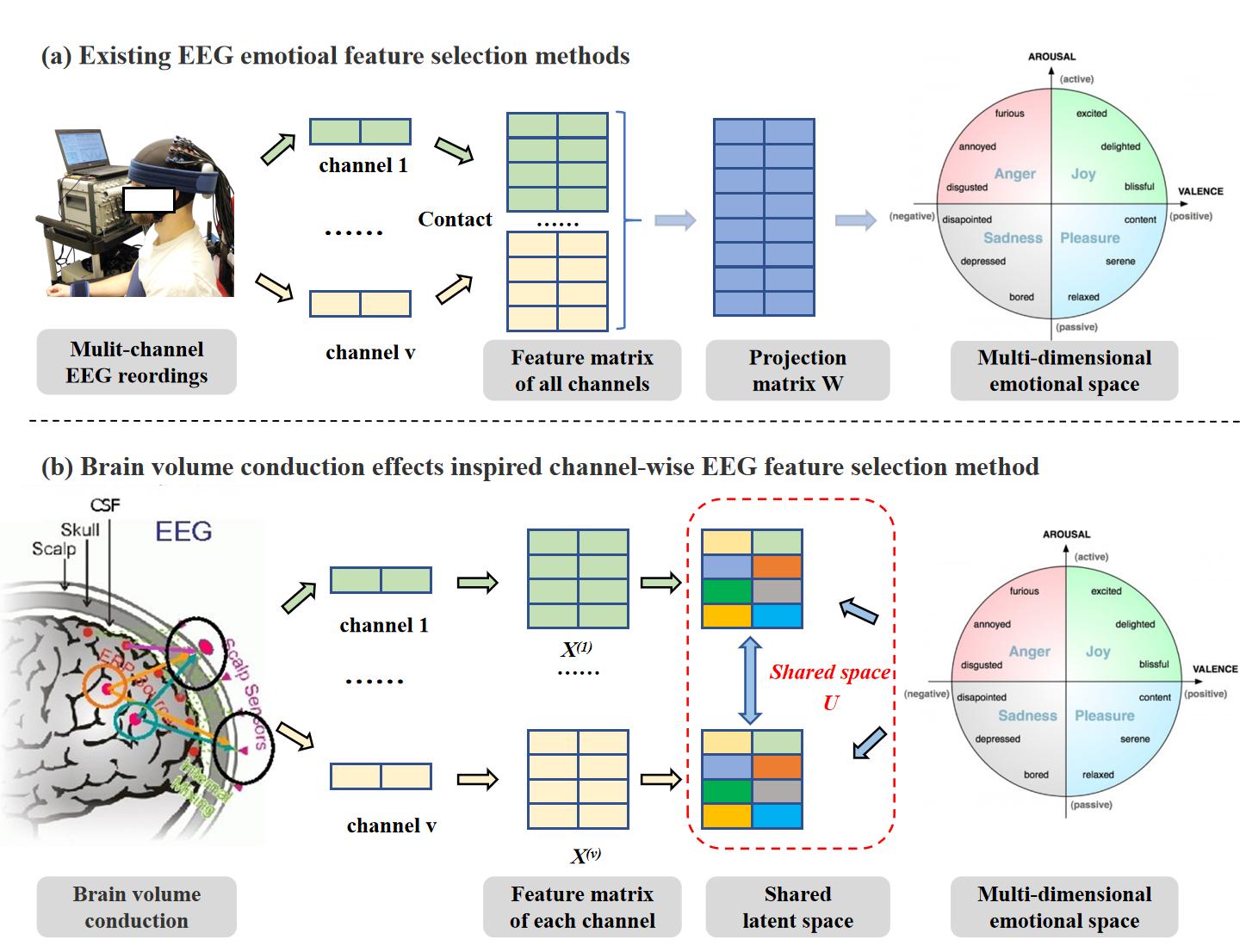}
\caption{(a) Current EEG feature selection methods intend to mine feature-label correlations via a projection subspace of original all EEG features in the emotional label space, which ignores the latent feature structure shared by all channels and all emotional dimensions; (b) To address the issue, a novel brain volume conduction effects inspired EEG feature selection method is proposed to construct a consensus latent space that aligns the multi-channel EEG feature spaces with multi-dimensional emotional label space.}
\label{Illustration}
\end{figure}

Advances in EEG signal acquisition technology have led to an increasing number of electrodes used in emotion recognition, enabling the extraction of larger EEG feature sets \cite{ozerdem2017emotion,Becker2020HRE}. However, the limited availability of EEG samples often results in high-dimensional feature spaces containing redundant, irrelevant, or noisy information, which can adversely affect the performance of emotion recognition systems \cite{wang2020emotion}. Feature selection offers a strategy to identify discriminative features while eliminating extraneous ones, preserving the neural information encoded in EEG signals and enhancing the interpretability and transparency of emotion recognition models \cite{GRMOR2021taffc,jenke2014taffc}.

EEG feature selection methods could be classified into three categories based on their criteria for evaluating feature subsets and searching for optimal solutions: filter, wrapper, and embedded methods \cite{saeys2007review}. Filter methods assess feature relevance using statistical properties of the data, but their performance often remains suboptimal regardless of the underlying learning algorithm \cite{zhang2019review}. Wrapper methods address this limitation by using the performance of a specific classifier as an evaluation criterion, frequently outperforming filter methods. However, they entail extensive computational trials and high costs \cite{li2017feature}. Recently, embedded methods have emerged as a promising alternative, integrating feature selection directly into the optimization process. Their effectiveness in EEG-based emotion recognition has been demonstrated in several studies \cite{liu2018electroencephalogram, xu2020fsorer, GRMOR2021taffc, xu2024embedded}.

The volume conduction effects in EEG refers to the phenomenon where electrical signals generated by neuronal activity propagate through the conductive media of the brain, skull, and scalp, leading to the spatial spread and mixing of potentials recorded at scalp electrodes as a superposition of contributions from multiple neural sources \cite{van1998volume}. Hence, the volume conduction properties introduce dependencies and independencies among multi-channel EEG features, indicating the presence of an intracranial latent structural hierarchy within multi-channel EEG data \cite{van1998volume,xu2020fsorer}. However, to guide the feature selection process, current EEG feature selection methods intend to mine feature-label correlations via a projection subspace of original EEG features in the emotion label space. The aforementioned approach ignores the impact of latent EEG feature structure on multi-dimensional emotion label correlations. Additionally, these methods assume that all channels have the same influence on emotional feature selection model construction. Nevertheless, a number of studies have found that electrodes in different brain regions exhibit distinct contributions to emotion modeling \cite{tao2020eeg,zheng2019identifying,GRMOR2021taffc}. 

To address these limitations, as illustrated in Fig.\ref{Illustration}, this paper proposes a novel channel-wise EEG feature selection (CWEFS) model for multi-dimensional emotion recognition, leveraging shared latent structure modeling and adaptive channel-weight learning. The model constructs a consensus latent space that aligns the multi-channel EEG feature spaces with multi-dimensional emotion label space, ensuring that similar EEG features are associated with similar emotional labels. Adaptive channel weighting is incorporated to automatically determine the relevance of individual channels during feature selection. Furthermore, graph-based manifold regularization learning is employed to preserve the local geometric relationships within both the EEG channel space and the multi-dimensional emotional label space, enhancing the accuracy of the consensus latent structure.

The contributions of this work are summarized as follows:
\begin{enumerate}
  \item[$\bullet$] Inspired by brain volume conduction effects, this study introduces a channel-wise EEG emotion feature selection method that integrates the feature selection process into a shared latent structure model, incorporating an adaptive channel-weight strategy. This approach could capture the differential influence of individual EEG channels on emotion recognition while modeling a common latent representation that aligns multi-channel EEG features with multi-dimensional emotion labels.
  \item[$\bullet$] To solve the optimization problem inherent in CWEFS, a straightforward yet efficient alternative optimization scheme is developed, ensuring convergence and enabling systematic identification of the optimal solution.
  \item[$\bullet$] To validate the effectiveness of CWEFS for the EEG based multi-dimensional emotion feature selection task, three benchmark datasets (DREAMER, DEAP, and HDED) comprising multi-channel EEG recordings and multi-dimensional emotion labels were employed. Experimental results demonstrate that the EEG feature subsets selected by CWEFS outperform those derived from nineteen popular feature selection methods, achieving superior multi-dimensional emotion recognition performance across six evaluation metrics.
\end{enumerate}

\section{Related Works}\label{RW}
\subsection{Notations and definitions}
This section offers a brief overview of the definitions for the norms and symbols employed throughout this work. Vectors  are denoted by lowercase boldface letters (e.g., $\mathbf{x}$, $\mathbf{y}$),  whereas matrices are represented by uppercase letters (e.g., $X$, $Y$). The transpose operation is denoted by a superscript uppercase $T$. The operator $\odot$  denotes the Hadamard product. The trace of a matrix is denoted as $\operatorname{Tr}$. 

%
%

The set $X = \left \lbrace X^{\left( 1 \right)}, X^{\left( 2 \right)}, ... , X^{\left( v \right)}, ... , X^{\left( ch \right)} \right \rbrace$ represents the multi-channel EEG feature data, where each element $X^{\left( v \right)} = [x_1, x_2, ... , x_{d_{v}}]^T$. Specifically, $X^{\left( v \right)} \in \mathbb{R}^{d_{v} \times n}$. $Y \in\{0,1\}^{k\times n}$ denotes the dimensional emotion label matrix. The matrices $U\in \mathbb{R}^{n \times k}$, $Q^{(v)}\in \mathbb{R}^{d_{v} \times k}$, and $M\in \mathbb{R}^{k \times k}$ are the shared common space across all EEG channels, the projection matrix for $X^{(v)}$, and the coefficient matrix for $Y$, respectively. The variables $d^{\left( v \right)}$, $n$, $ch$, and $k$ denote the number of features for each channel, the number of instances, the number of channels, and the number of label dimensions, respectively.

\subsection{EEG feature selection methods}

Current feature selection methods for brain-computer interfaces are primarily categorized into three classes: filter-based, wrapper-based, and embedded approaches\cite{saeys2007review}. Filter-based methods independently assess feature relevance using statistical or information-theoretic metrics, such as Pearson correlation coefficients (PCC), minimal redundancy maximal relevance (mRMR)\cite{wang2011eeg,atkinson2016improving}, and information gain\cite{chen2015electroencephalogram}. While computationally efficient, these methods may overlook feature combinations optimized for specific classifiers due to their independence from the model-building process\cite{zhang2019review}. Wrapper-based methods guide feature search through classifier performance feedback, exemplified by ant colony optimization and particle swarm optimization algorithms\cite{dorigo1997ant,he2020strengthen}. However, their high computational complexity limits real-time applications\cite{saeys2007review}.

Embedded methods integrate feature selection into model training, simultaneously achieving feature importance evaluation and model construction by optimizing objective functions, with regularization techniques automatically prioritizing critical features. Among these, least squares regression-based embedded methods are widely adopted due to their statistical theoretical rigor\cite{tang2014feature}. Several least squares regression-based embedded frameworks have been proposed or implemented for addressing the EEG feature selection issue, including: robust feature selection (RFS)\cite{nie2010RFS}, feature selection with orthogonal regression (FSOR) \cite{xu2020fsorer},global redundancy minimization in orthogonal regression (GRMOR)\cite{GRMOR2021taffc}, EEG feature selection for multi-dimension emotion recognition (EFSMDER)\cite{xu2024embedded}. 

\begin{figure*}[!t]
\centering
\includegraphics[width=0.732\textwidth]{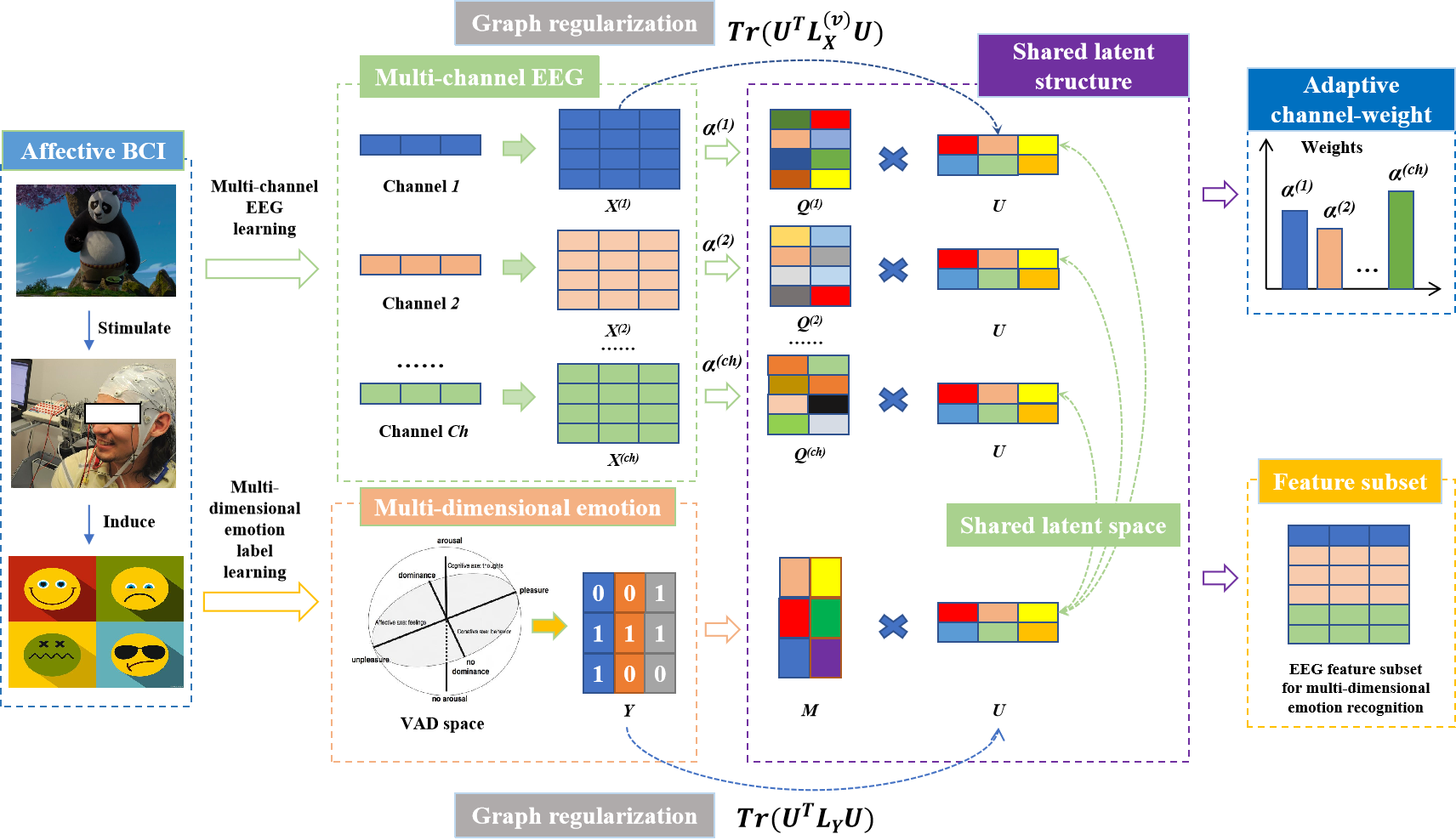}
\caption{The proposed CWEFS framework concludes the following two sections: (a) shared latent structure learning with adaptive channel-weight; (b) graph regularization learning.}
\label{Framework_efsmder}
\end{figure*}

\section{Problem formulation}\label{Pf}
CWEFS comprises shared latent structure learning with adaptive channel-weight and graph regularization learning. Subsequent subsections formally define their mathematical expressions.

\subsection{Shared latent structure learning with adaptive channel-weight}
The conventional non-negative matrix factorization model decomposes a matrix $X$ into two non-negative matrices such that $X \approx Q U^T$, with its cost function formulated as: 
\begin{equation}
\begin{gathered}
\min _{Q, U} \left\|X-Q U^T\right\|_F^2  \text { s.t. } Q \geq 0, U \geq 0 \\
\end{gathered}
\end{equation}
where $U$ represents the latent structure of the matrix $X$, and $Q$ denotes the coefficient matrix.

Extending this framework, the $v$-channel EEG feature matrix $X^{(v)}$ and multi-dimensional emotion label matrix $Y$ are respectively factorized as $X^{(v)} \approx Q^{(v)} U^T$ and $Y \approx M U^T$. By leveraging the premise that neurophysiologically similar EEG features correlate with analogous emotional states, we construct a shared latent structure space through the common matrix $U$. This unified representation establishes dependencies between EEG features and multi-dimensional emotion labels, formally expressed as:
\begin{equation}
\begin{gathered}
\label{equ_framework_lls}
\min _{Q^{(v)}, M, U} \left\|X^{(v)}-Q^{(v)} U^T\right\|_F^2+\lambda\left\|Y-M U^T\right\|_F^2 \\
\text { s.t. } Q^{(v)} \geq 0, M \geq 0, U \geq 0 \\
\end{gathered}
\end{equation}

Building upon empirical evidence \cite{tao2020eeg,zheng2019identifying,GRMOR2021taffc} that EEG channels exhibit heterogeneous contributions to emotion modeling, we develop an adaptive weighting mechanism to allocate channel-specific coefficients $\alpha^{(v)}$ ($v=1, \ldots, ch$). Inspired by brain volume conduction effects, to holistically capture cross-channel dependencies, a shared latent feature structure $U$ is maintained across all EEG channels. In summary, Eq.~\eqref{equ_framework_lls} can be changed to:
\begin{equation}
\begin{gathered}
\label{equ_framework_F}
\min _{\substack{Q^{(v)}, M,\\ \alpha^{(v)}, U}} \sum_{v=1}^{ch}\left(\alpha^{(v)}\right)^\gamma\left(\left\|X^{(v)}-Q^{(v)} U^T\right\|_F^2+\lambda\left\|Y-M U^T\right\|_F^2\right) \\
\text { s.t. } Q^{(v)} \geq 0, M \geq 0, U \geq 0, \\
\sum_{v=1}^{ch} \alpha^{(v)}=1, 0 \leq \alpha^{(v)} \leq 1, v=1, \ldots, ch
\end{gathered}
\end{equation}

\subsection{Graph regularization learning}
Drawing upon spectral graph theory \cite{jian2016multi}, we incorporate dual manifold regularization to maintain consistency of the local geometric structures. Specifically, the term $\operatorname{Tr}\left(U^T L_Y U\right)$ enforces geometric consistency between the shared latent space $U$ and the affective label space $Y$, while $\operatorname{Tr}\left(U^T L_X^{(v)} U\right)$ maintains topological fidelity within individual EEG channels. The composite graph regularization term is then formulated as:
\begin{equation}
\begin{gathered}
\label{equ_framework_C}
\min _{U}  \operatorname{Tr}\left(U^T L_Y U\right) +\beta \operatorname{Tr}\left(U^T L_X^{(v)} U\right) \text { s.t. } U \geq 0 \\
\end{gathered}
\end{equation}
where the graph Laplacian matrix $L_Y\in {\mathbb{R}^{\text{n}\times n}}$ for $Y$ is indicated as $L_{Y}=G_{Y}-S_{Y}$. Here, $S_{Y}$ represents the affinity graph of $Y$, and $G_{Y}$ denotes the diagonal matrix with $G_{Y}(i,i)=\sum_{j=1}^{n} S_{Y}(i,j)$. Similarly, the graph Laplacian matrix $L_X^{(v)}\in {\mathbb{R}^{\text{n}\times n}}$ for $X^{(v)}$ is defined by $L_{X}^{(v)}=G_{X}^{(v)}-S_{X}^{(v)}$. The parameter $\beta$ denotes a tradeoff coefficient.

A heat kernel is used to generate the affinity graphs $S_{Y}$ and $S_{X}^{(v)}$. The similarity between two instances, $\bm{x}_{.i}^{(v)}$ and $\bm{x}_{.j}^{(v)}$, is denoted by the element $S_{X}^{(v)}(i,j)$, which is defined as follows:
\begin{equation}
S_{X}^{(v)}(i,j)=\left\{\begin{array}{lc}
\exp \left(-\frac{\left\|\bm{x}_{.i}^{(v)}-\bm{x}_{.j}^{(v)}\right\|^{2}}{\sigma^{2}}\right) &\bm{x}_{.i}^{(v)} \in \mathcal{N}_{q}\left(\bm{x}_{.j}^{(v)}\right) \text { or } \\ &\bm{x}_{.j}^{(v)} \in \mathcal{N}_{q}\left(\bm{x}_{.i}^{(v)}\right) \\
0 & \text { otherwise }
\end{array}\right.
\end{equation}

The symbol $\sigma$ denotes the graph construction parameter, while $\mathcal{N}_{p}\left(\bm{x}_{.j}^{(v)}\right)$ represents the set of the $q$ nearest neighbors of the sample $\bm{x}_{.j}^{(v)}$. Referring to the study\cite{jian2016multi}, the value of $\sigma$ is set to 1.

Furthermore, the similarity between two labels $\bm{y}_{.i}$ and $\bm{y}_{.j}$ is denoted by the element $S_{Y}(i,j)$. $S_{Y}(i,j)$, which is defined as follows:
\begin{equation}
S_{Y}(i,j)=\left\{\begin{array}{lc}
\exp \left(-\frac{\left\|\bm{y}_{.i}-\bm{y}_{.j}\right\|^{2}}{\sigma^{2}}\right) &\bm{y}_{.i} \in \mathcal{N}_{q}\left(\bm{y}_{.j}\right) \text { or } \\ & \bm{y}_{.j} \in \mathcal{N}_{q}\left(\bm{y}_{.i}\right) \\
0 & \text { otherwise }
\end{array}\right.
\end{equation}
where $\mathcal{N}_{p}\left(\bm{y}_{.j}\right)$ denotes the set of the $q$ nearest neighbors of the label $\bm{y}_{.j}$.

\subsection{The final objective function of CWEFS}
By integrating Eq.~\eqref{equ_framework_F} and Eq.~\eqref{equ_framework_C} and imposing the $l_{2,1}$-norm regularization on $Q^{(v)}$, we formulate the channel-wise EEG feature selection framework as:
\begin{equation}
\begin{gathered}
\label{equ_framework}
\min _{\substack{Q^{(v)}, M,\\ \alpha^{(v)}, U}} \sum_{v=1}^{ch}\left(\alpha^{(v)}\right)^\gamma\left(\left\|X^{(v)}-Q^{(v)} U^T\right\|_F^2+\lambda\left\|Y-M U^T\right\|_F^2\right. \\
\left.+\eta \operatorname{Tr}\left(U^T L_Y U\right) +\beta \operatorname{Tr}\left(U^T L_X^{(v)} U\right)  +\delta\left\|Q^{(v)}\right\|_{2,1}\right)\\
\text { s.t. } Q^{(v)} \geq 0, M \geq 0, U \geq 0, \\
\sum_{v=1}^{ch} \alpha^{(v)}=1, 0 \leq \alpha^{(v)} \leq 1, v=1, \ldots, ch
\end{gathered}
\end{equation}
where $\lambda$, $\beta$, $\eta$, $\gamma$, and $\delta$ are tradeoff parameters. The architectural workflow of CWEFS is formally depicted in Fig.~\ref{Framework_efsmder}. 

\section{Optimization Strategy} \label{Optimization Strategy}
An alternating optimization scheme is employed to compute closed-form solutions for the variables ($Q^{(v)}$, $U$, $M$, and $\alpha^{(v)}$) within the unified framework of Eq.~\eqref{equ_framework}. The numerical procedure is implemented through the following steps:

\subsection{Update $Q^{(v)}$ by fixing $U$, $M$, and $\alpha^{(v)}$}
Upon fixing $U$, $M$, and $\alpha^{(v)}$, we eliminate extraneous components and formally incorporate a Lagrange multiplier $\mathbf{\Psi}$ to enforce the non-negativity constraint $Q^{(v)} \geq 0$, yielding:
\begin{equation}
\begin{aligned}
\label{equ_q_l}
\mathcal{L}\left(Q^{(v)}\right) = &\left(\alpha^{(v)}\right)^\gamma\left(\left\|X^{(v)}-Q^{(v)} U^T\right\|_F^2+\delta\left\|Q^{(v)}\right\|_{2,1}\right)\\
 &+\operatorname{Tr}\left(\mathbf{\Psi}^T Q^{(v)}\right)
\end{aligned}
\end{equation}

The partial derivative of $\mathcal{L}\left(Q^{(v)}\right)$ with respect to $Q^{(v)}$ is:
\begin{equation}
\begin{aligned}
\label{equ_qv_pd}
\frac{\partial \mathcal{L}\left(Q^{(v)}\right)}{\partial Q^{(v)}} = &\left(\alpha^{(v)}\right)^\gamma\left(2 Q^{(v)} U^T U - 2 X^{(v)} U +2 \delta D^{(v)} Q^{(v)}\right)\\
 &+ \mathbf{\Psi}
\end{aligned}
\end{equation}
where $D^{(v)}$ denotes a diagonal matrix whose diagonal entries are defined as $D_{i i}^{(v)}=\frac{1}{2 \sqrt{{Q^{(v)}_i}^T Q^{(v)}_i+\epsilon}}(\epsilon \rightarrow 0)$.

Under the Karush-Kuhn-Tucker (KKT) complementary conditions, $Q^{(v)}$ can be presented as:
\begin{equation}
\label{equ_qv_sol}
Q^{(v)} \leftarrow Q^{(v)} \odot \frac{X^{(v)} U}{Q^{(v)} U^T U + \delta D^{(v)} Q^{(v)}}
\end{equation}

\subsection{Update $U$ by fixing $Q^{(v)}$, $M$, and $\alpha^{(v)}$}
When $Q^{(v)}$, $M$, and $\alpha^{(v)}$ are fixed, the Lagrangian function is constructed by introducing a Lagrange multiplier $\mathbf{\Theta}$ associated with the constraint $U \geq 0$, yielding:
\begin{equation}
\begin{aligned}
\label{equ_u_pd}
\mathcal{L}\left(U\right)=\sum_{v=1}^{ch}\left(\alpha^{(v)}\right)^\gamma\left(\left\|X^{(v)}-Q^{(v)} U^T\right\|_F^2+\lambda\left\|Y-M U^T\right\|_F^2\right. \\
\left.+\eta \operatorname{Tr}\left(U^T L_Y U\right) +\beta \operatorname{Tr}\left(U^T L_X^{(v)} U\right)\right) +\operatorname{Tr}\left(\mathbf{\Theta}^T U\right)\\
\end{aligned}
\end{equation}

The partial derivative of $\mathcal{L}\left(U\right)$ with respect to $U$ is:
\begin{equation}
\begin{aligned}
\frac{\partial \mathcal{L}\left(U\right)}{\partial U}=\sum_{v=1}^{ch}\left(\alpha^{(v)}\right)^\gamma\left(2 U {Q^{{(v)}}}^T Q^{(v)} - 2 {X^{(v)}}^T Q^{(v)} \right.  \\
\left. + 2 \lambda U M^T M - 2 \lambda Y^T M + 2 \eta L_Y U + 2 \beta L_X^{(v)} U\right) + \mathbf{\Theta}\\
\end{aligned}
\end{equation}

Under the KKT complementary condition specified by $\mathbf{\Theta}_{ij}U_{ij} = 0$, $U$ can be presented as:
\begin{equation}
\label{equ_u_sol}
U \leftarrow U \odot \frac{\sum_{v=1}^{ch}\left(\alpha^{(v)}\right)^\gamma\left({X^{(v)}}^T Q^{(v)}+\lambda Y^T M \right) }{\sum_{v=1}^{ch}\left(\alpha^{(v)}\right)^\gamma\left(Z\right) }
\end{equation}
where $Z = U {Q^{(v)}}^T Q^{(v)}+\beta L_X^{(v)} U+ \eta L_Y U+ \lambda U M^T M $.

\subsection{Update $M$ by fixing $Q^{(v)}$, $U$, and $\alpha^{(v)}$}
When $Q^{(v)}$, $U$, and $\alpha^{(v)}$ are fixed, the Lagrangian function is constructed by introducing a Lagrange multiplier $\mathbf{\Phi}$ associated with the constraint $M \geq 0$, yielding:
\begin{equation}
\begin{aligned}
\label{equ_m_pd}
\mathcal{L}\left(M\right)=\sum_{v=1}^{ch}\left(\alpha^{(v)}\right)^\gamma\left(\lambda\left\|Y-M U^T\right\|_F^2\right) +\operatorname{Tr}\left(\mathbf{\Phi}^T M\right) \\
\end{aligned}
\end{equation}

The partial derivative of $\mathcal{L}\left(M\right)$ with respect to $M$ is:
\begin{equation}
\begin{aligned}
\frac{\partial \mathcal{L}\left(M\right)}{\partial M}=\sum_{v=1}^{ch}\left(\alpha^{(v)}\right)^\gamma\left(2 \lambda M U^T U - 2 \lambda Y U \right) + \mathbf{\Phi}  \\
\end{aligned}
\end{equation}

Under the KKT complementary condition specified by $\mathbf{\Phi}_{ij}M_{ij} = 0$, $M$ can be presented as:
\begin{equation}
\label{equ_m_sol}
M \leftarrow M \odot \frac{\sum_{v=1}^{ch}\left(\alpha^{(v)}\right)^\gamma\left(Y U \right) }{\sum_{v=1}^{ch}\left(\alpha^{(v)}\right)^\gamma\left(M U^T U\right) }
\end{equation}

\subsection{Update $\alpha^{(v)}$ by fixing $Q^{(v)}$, $U$, and $M$}
When all other variables are held constant, we have:
\begin{equation}
\begin{aligned}
e^{(v)} = &\left\|X^{(v)}-Q^{(v)} U^T\right\|_F^2+\lambda\left\|Y-M U^T\right\|_F^2 +\delta\left\|Q^{(v)}\right\|_{2,1}\\
&+\eta \operatorname{Tr}\left(U^T L_Y U\right) +\beta \operatorname{Tr}\left(U^T L_X^{(v)} U\right) 
\end{aligned}
\end{equation}

Then, the optimization problem of $\alpha^{(v)}$ is changed to:
\begin{equation}
\min _{\alpha^{(v)}} \sum_{v=1}^{ch}\left(\alpha^{(v)}\right)^\gamma e^{(v)} \text {, s.t. } \sum_{v=1}^{ch} \alpha^{(v)}=1,0 \leq \alpha^{(v)} \leq 1
\end{equation}

The Lagrangian function is constructed through the introduction of a Lagrange multiplier $\mathbf{\xi}$ corresponding to the constraint $\sum_{v=1}^{ch} \alpha^{(v)}=1$, resulting in:
\begin{equation}
\label{equ_alpha_obj}
\mathcal{L}\left(\alpha^{(v)}\right)=\sum_{v=1}^{ch} \left(\alpha^{(v)}\right)^\gamma e^{(v)}-\mathbf{\xi} \left(\sum_{v=1}^{ch} \alpha^{(v)}-1\right)
\end{equation}

The partial derivative of $\mathcal{L}\left(\alpha^{(v)}\right)$ with respect to $\alpha^{(v)}$ is:
\begin{equation}
\frac{\partial \mathcal{L}\left(\alpha^{(v)}\right)}{\partial \alpha^{(v)}}=\gamma\left(\alpha^{(v)}\right)^{\gamma-1} e^{(v)}-\xi
\end{equation}

Setting the partial derivative of $\frac{\partial \mathcal{L}\left(\alpha^{(v)}\right)}{\partial \alpha^{(v)}}$ to zero yields:
\begin{equation}
\label{equ_alpha_ori}
\alpha^{(v)}=\left(\frac{\xi}{\gamma e^{(v)}}\right)^{1 /(\gamma-1)}
\end{equation}

By virtue of the constraint $\sum_{v=1}^{ch} \alpha^{(v)}=1$, Eq.~\eqref{equ_alpha_ori} can be reformulated in the following form:
\begin{equation}
\label{equ_alpha_sol}
\alpha^{(v)}=\frac{\left(e^{(v)}\right)^{\frac{1}{1-\gamma}}}{\sum_{v=1}^{ch}\left(e^{(v)}\right)^{\frac{1}{1-\gamma}}}
\end{equation}

\begin{algorithm}[h]
\caption{Iterative algorithm of CWEFS}
\label{CWEFS}
\begin{algorithmic}[1]
\Require

1) multi-channel EEG feature data $X$;

2) multi-dimensional emotional matrix $Y\in {\mathbb{R}^{k\times n}}$;

3) tradeoff parameters $\lambda$, $\beta$, $\eta$, $\mu$, and $\delta$.
\Ensure Return ranked EEG features.
\State Initial $Q^{(v)}$, $U$, and $M$ randomly. $\alpha^{(v)} = 1/ch$.
\Repeat
\State Update $Q^{(v)}$ via Eq.~\eqref{equ_qv_sol};
\State Update $U$ via Eq.~\eqref{equ_u_sol};
\State Update $M$ via Eq.~\eqref{equ_m_sol};
\State Update $\alpha^{(v)}$ via Eq.~\eqref{equ_alpha_sol};
\Until{Convergence;}
\State \Return $Q^{(v)}$ for EEG feature selection.
\State Sort the EEG features by $ \|Q^{(v)}_{i}\|_{2}$;
\end{algorithmic}
\end{algorithm}

Algorithm~\ref{CWEFS} details the precise optimization procedure for Eq.~\eqref{equ_framework}. The significance of each feature within the emotion recognition task is subsequently quantified via $Q^{(v)}$. Through this process, the EEG feature subset is chosen.

\section{Experiments} \label{Experimental Details}

\subsection{Dataset description}
A comprehensive experimental evaluation was performed across three benchmark EEG datasets with multi-dimensional emotional annotations to systematically evaluate the efficacy of CWEFS, encompassing DREAMER~\cite{DREAMER2018jbhi}, DEAP~\cite{koelstra2011deap}, and HDED~\cite{GRMOR2021taffc}. The datasets implemented the valence-arousal-dominance (VAD) paradigm to characterize human affective states. Throughout multimedia stimulation protocols, synchronized EEG recordings were acquired. Detailed experimental configuration parameters and acquisition protocols are documented in the original studies~\cite{DREAMER2018jbhi, koelstra2011deap, GRMOR2021taffc}.

\subsection{EEG feature extraction}
The experimental pipeline implemented a band-pass filtering operation (1-50 Hz cutoff frequency) to eliminate noise from raw EEG recordings. Ocular and myogenic artifacts were subsequently mitigated through independent component analysis. Notably, the EEG feature extraction process was conducted on entire trial epochs as unified samples. Specifically, trial segmentation was intentionally avoided to preserve the original sample integrity and prevent artificial inflation of the experimental dataset size.

Building upon foundational EEG affective computing methodologies~\cite{jenke2014taffc,xu2020fsorer}, thirteen neurophysiological feature types were systematically derived for multi-dimensional emotion recognition: C0 complexity, non-stationary index, DASM, higher-order crossing, spectral entropy, rational asymmetry, shannon entropy, DE, absolute power, the absolute power ratio of the theta band to the beta band, the amplitude of the Hilbert transform of intrinsic mode functions, the instantaneous phase of the Hilbert transform of intrinsic mode functions, and function connectivity. Detailed mathematical formalisms and physiological interpretations of the feature types are documented in~\cite{Duan2013DE,jenke2014taffc,GRMOR2021taffc}. In summary, the feature dimensions of DREAMER, DEAP, and HDED are 651, 1756, and 7565, respectively.


\subsection{Experimental setup}
To comprehensively assess the performance of CWEFS in the multi-dimensional affective computing task, nineteen state-of-the-art feature selection methods were included in the comparative analysis. These methods include:

(1) Ten single-label feature selection approaches: ReliefF \cite{kononenko1994ReliefF}, PCC \cite{ng1997feature}, CMIM \cite{fleuret2004CMIM}, mRMR \cite{ding2005mRMR}, RFS \cite{nie2010RFS}, RPMFS \cite{cai2013sfs}, ESFS \cite{chen2017embedded}, FSOR \cite{xu2020fsorer}, SDFS \cite{wang2020discriminative}, and GRMOR \cite{GRMOR2021taffc}.

(2) Five multi-label feature selection methods: PMU \cite{lee2013feature}, SCLS \cite{lee2017scls}, MFS\_MCDM \cite{hashemi2020mfs}, MGFS \cite{hashemi2020mgfs}, and EFSMDER \cite{xu2024embedded}.

(3) Four advanced multi-view multi-label feature selection methods: MSFS \cite{zhang2020multi}, DHLI \cite{hao2024double}, UGRFS \cite{hao2025uncertainty}, and EF$^2$FS \cite{hao2025embedded}. 

The EEG recordings were stratified into dichotomous classes (low/high) according to self-assessed emotion dimensional scores, with an empirically established classification threshold of 5. Multi-label k-nearest neighbors (ML-KNN)~\cite{ZHANG2007MLKNN} was adopted as the base classifier, where the neighborhood size and smoothing parameters were configured to $k=10$ and $s=1$, respectively. The dataset was partitioned through random allocation, with 80\% of participants assigned to the training set and the remaining 20\% retained as the test set. A cross-subject validation paradigm was employed, complemented by 50 independent trials to eliminate sampling bias. The mean performance metric across all trials served as the definitive evaluation criterion for affective computing efficacy. Six evaluation metrics were employed to assess multi-dimensional emotion recognition performance: (1) label-based metrics: macro-F1 (MA) and micro-F1 (MI); (2) instance-based metrics: average precision (AP), coverage (CV), ranking loss (RL), and hamming loss (HL). The mathematical definitions of the metrics are detailed in~\cite{zhang2019manifold}.

\begin{figure}[!t]
\centering
\includegraphics[width=0.479\textwidth]{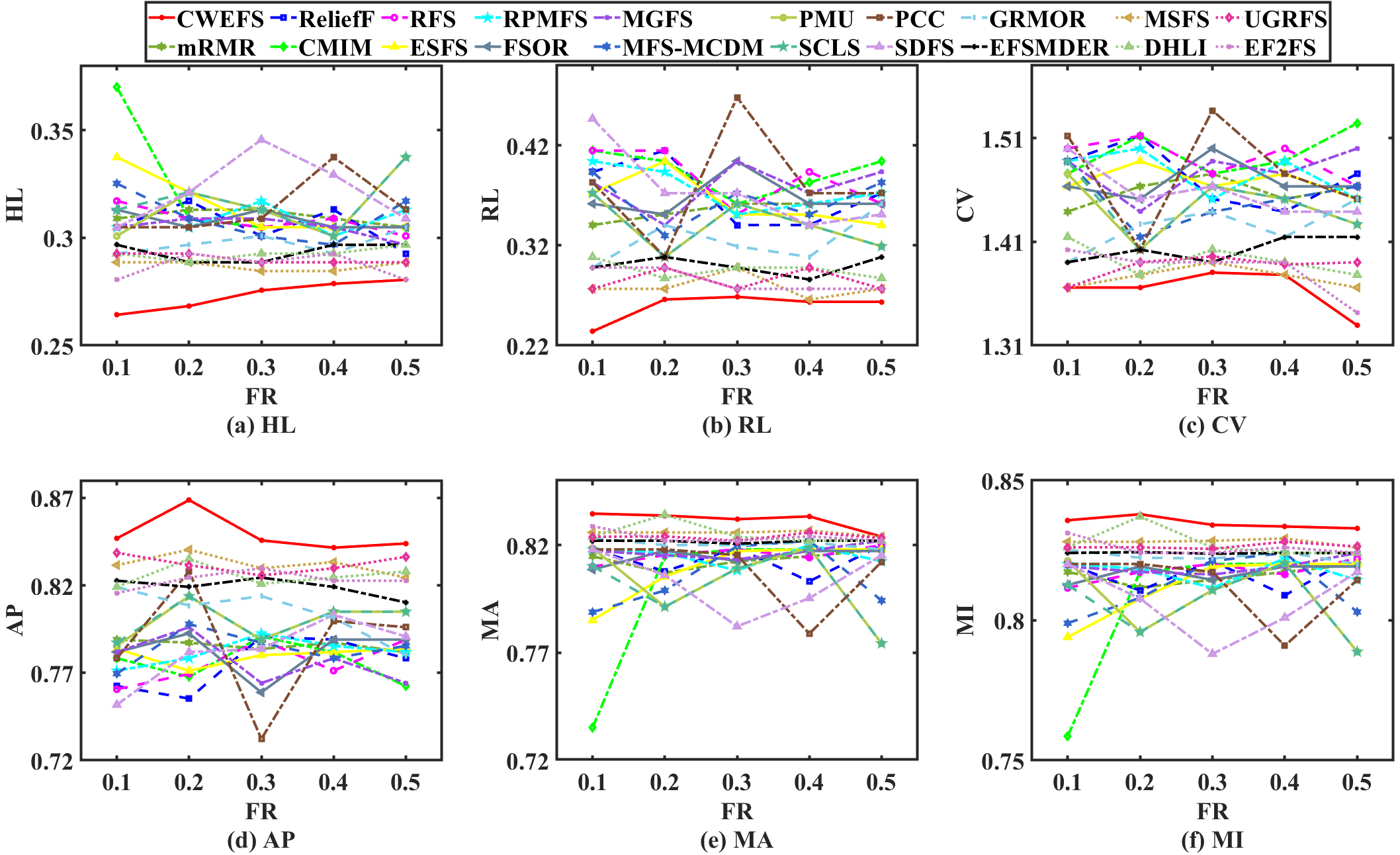}
\caption{Multi-dimensional emotion recognition performance of various feature ratios (FR) on DREAMER.}\label{Results_index_dreamer}
\end{figure}
\begin{figure}[!t]
\centering
\includegraphics[width=0.479\textwidth]{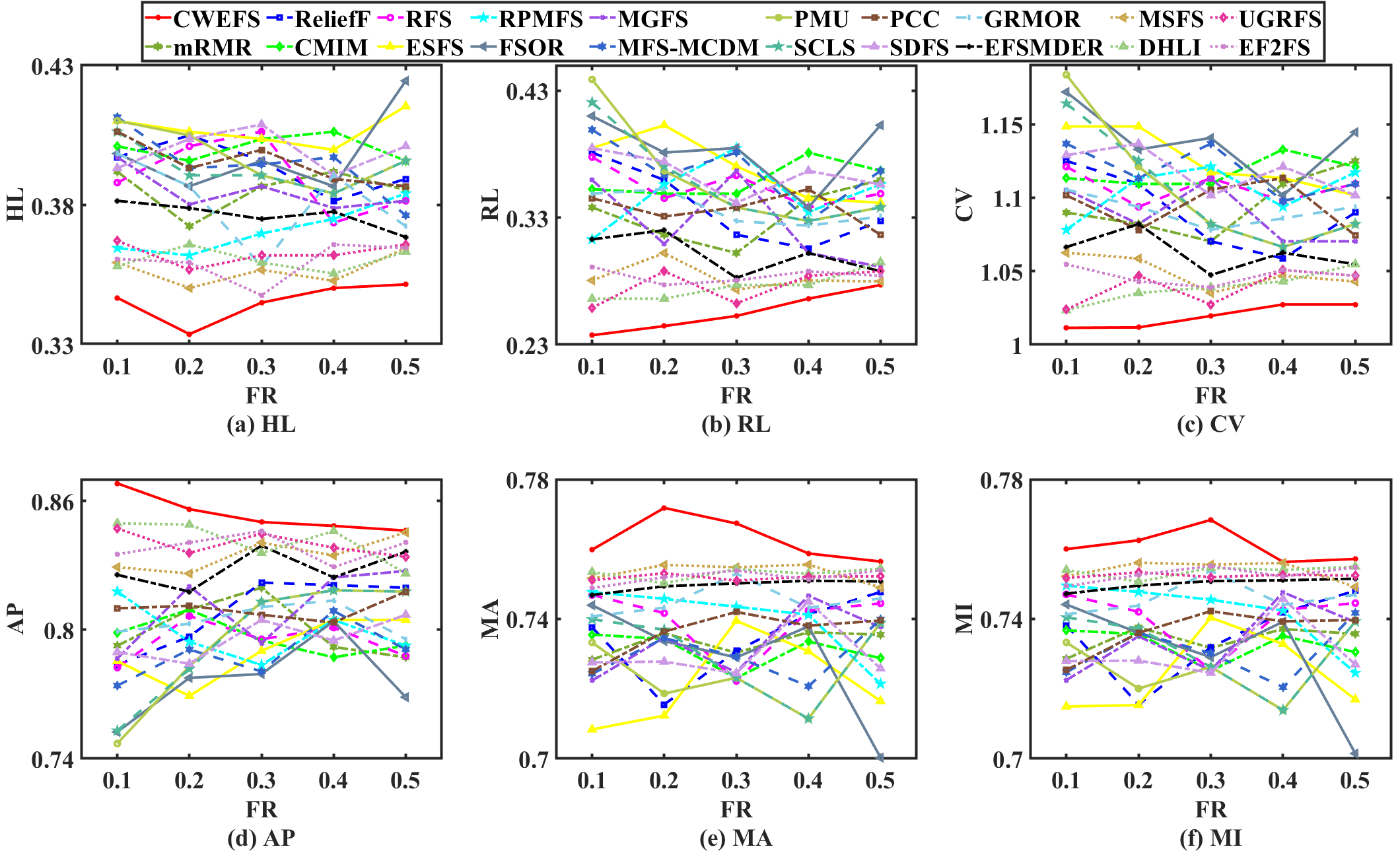}
\caption{Multi-dimensional emotion recognition performance of various feature ratios on DEAP.}\label{Results_index_deap}
\end{figure}
\begin{figure}[!t]
\centering
\includegraphics[width=0.479\textwidth]{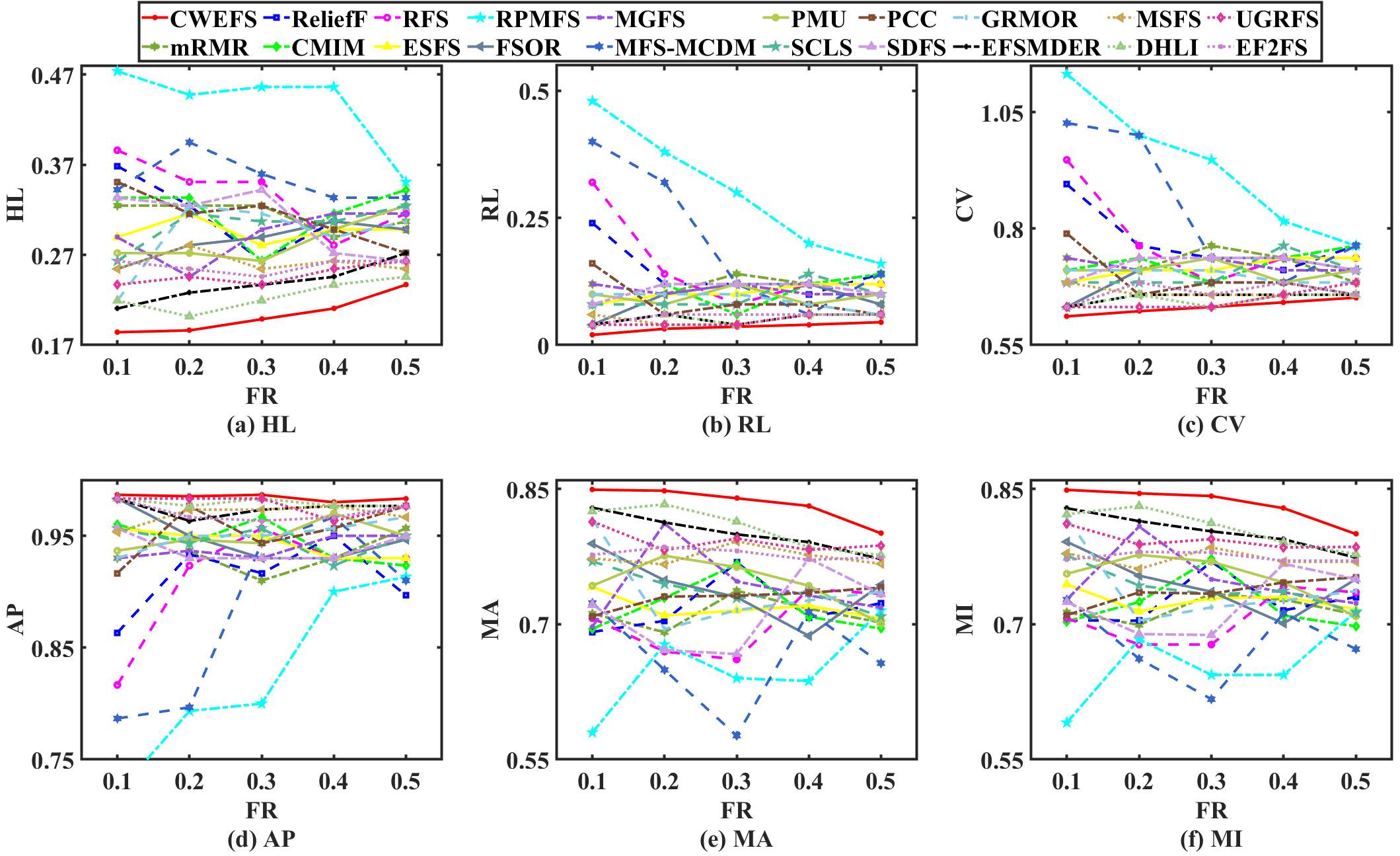}
\caption{Multi-dimensional emotion recognition performance of various feature ratios on HDED.}\label{Results_index_hded}
\end{figure}

\begin{table}[!t]\small
\begin{center}
{
\begin{tabular}{lcc}
\hline\hline
Evaluation metric     & $F_{F}$             & Critical value  \\ \hline
Ranking loss          & 6.919              & \multicolumn{1}{c}{\multirow{7}*{$\approx$ 1.867}}             \\
Coverage              & 7.994             &            \\
Hamming loss          & 7.754             &            \\
Average precision     & 9.303             &           \\
Macro-F1              & 7.396            &            \\
Micro-F1              & 7.500             &            \\
\hline\hline
\end{tabular}}
\caption{Friedman test results.}\label{tab:friedman}
\end{center}
\end{table}

\begin{table}[!t]\small
\begin{center}
{
\begin{tabular}{lcc||cc||cc}
\hline\hline
\multicolumn{1}{l}{\multirow{2}*{Methods}}          & \multicolumn{2}{c}{\multirow{1}*{DREAMER}}   & \multicolumn{2}{c}{\multirow{1}*{DEAP}}  & \multicolumn{2}{c}{\multirow{1}*{HDED}} \\  \cline{2-7}
                         & HL $\downarrow$             & AP$\uparrow$  & HL $\downarrow$             & AP$\uparrow$ & HL $\downarrow$             & AP$\uparrow$     \\\hline
w/o ACWL                 &0.30      &0.80      &0.36      &0.83   &0.25       &0.96   \\
w/o GMRL\_Y              &0.30      &0.79       &0.37      &0.82   &0.28       &0.94      \\
w/o GMRL\_X              &0.29      &0.81       &0.36      &0.84    &0.23       &0.97    \\
\textbf{CWEFS}           & \textbf{0.28}             & \textbf{0.83}   & \textbf{0.35}  & \textbf{0.85} &\textbf{0.20}     &\textbf{0.98}  \\ \hline\hline
\end{tabular}}
\caption{The results of ablation experiments on the performance index average precision (w/o, ACWL, and GMRL denote without, adaptive channel-weight learning, and graph-based manifold regularization learning, respectively).}\label{tab:ab}
\end{center}
\end{table}
\subsection{Performance comparison}
The self-assessed VAD dimensions were stratified into dichotomous classes (high/low), thereby reformulating the multi-dimensional emotion recognition task as a multi-label learning problem. A comparative evaluation of feature selection ratios is visualized in Fig.~\ref{Results_index_dreamer}, Fig.~\ref{Results_index_deap}, and Fig.~\ref{Results_index_hded}. Horizontal axes denote the proportion of chosen features, and vertical axes quantify multi-label classification performance metrics. CWEFS is highlighted in a red line across sub-figures. 

As shown in Fig. 3(a-c), Fig. 4(a-c), Fig. 5(a-c), enhanced multi-dimensional affective computing performance correlates with lower values of three evaluation metrics: HL, CV, and RL. Conversely, as demonstrated in Fig. 3(d-f), Fig. 4(d-f), and Fig. 5(d-f), superior performance is indicated by elevated values of three complementary metrics: AP, MA, and MI. Across all EEG feature selection ratios, CWEFS systematically attains extremal values (minima for HL/CV/RL and maxima for AP/MA/MI), outperforming nineteen benchmark feature selection methods. Collectively, the experimental evidence from Fig. 3-5 substantiates that EEG feature subsets chosen by CWEFS achieve peak recognition efficacy across all evaluation criteria. Furthermore, the Friedman test was employed to statistically validate significant performance disparities among the twenty methods, as quantified in Table~\ref{tab:friedman}. The null hypothesis was rejected, indicating significant differences in these twenty methods.

\subsection{Ablation experiment}
To evaluate the contributions of individual modules within the CWEFS framework, we conducted controlled ablation studies. CWEFS comprises three critical components, which were systematically deactivated in sequence. As summarized in Table~\ref{tab:ab}, the adaptive channel-weight learning module plays a pivotal role in characterizing the varying impacts of individual EEG channels on feature selection model construction. The remaining modules serve to preserve the local geometric structures within both the original EEG feature space and the multi-dimensional affective label space.

\begin{figure}[!t]
\centering
\subfigure[$\gamma$]{\label{bar3_gamma}\includegraphics[width=0.14\textwidth]{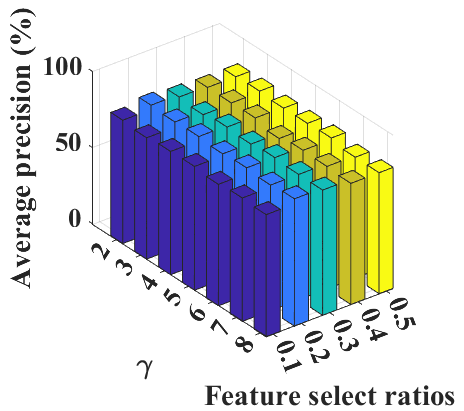}}
\hspace{0.0cm}
\subfigure[$\lambda$]{\label{bar3_lambda}\includegraphics[width=0.14\textwidth]{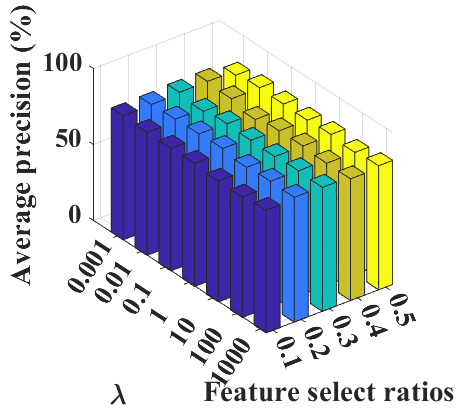}}
\hspace{0.0cm}
\subfigure[$\eta$]{\label{bar3_eta}\includegraphics[width=0.14\textwidth]{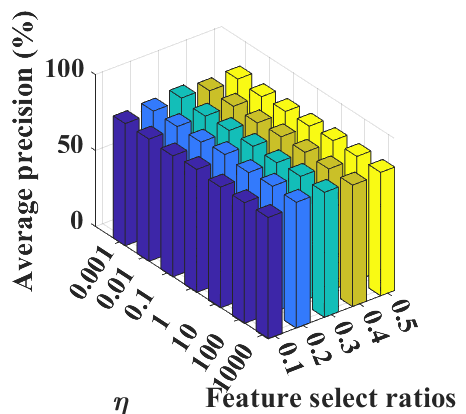}}
\hspace{0.0cm}
\subfigure[$\beta$]{\label{bar3_beta}\includegraphics[width=0.14\textwidth]{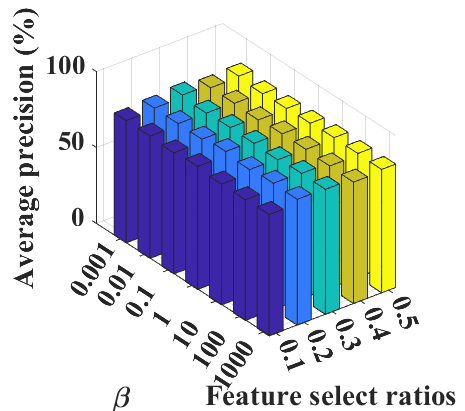}}
\hspace{0.0cm}
\subfigure[$\delta$]{\label{bar3_delta}\includegraphics[width=0.14\textwidth]{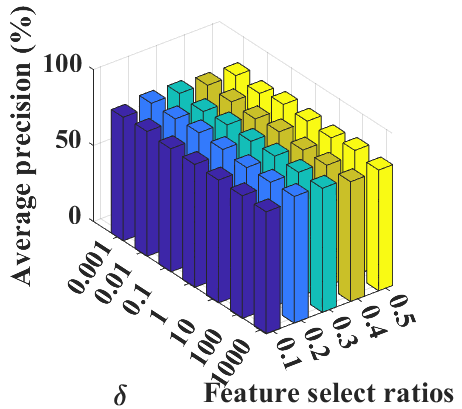}}
\caption{The parameter sensitivity of CWEFS on DEAP.}
\label{Results_bar3_deap}
\end{figure}
\begin{figure}[!t]
\centering
\subfigure[DREAMER]{\label{Conv_dreamer}\includegraphics[width=0.151\textwidth]{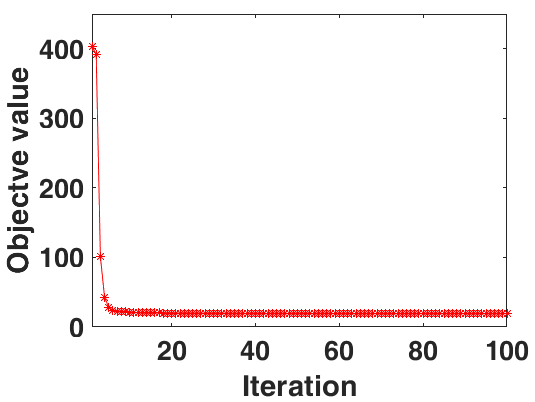}}
\hspace{0.0cm}
\subfigure[DEAP]{\label{Conv_deap}\includegraphics[width=0.151\textwidth]{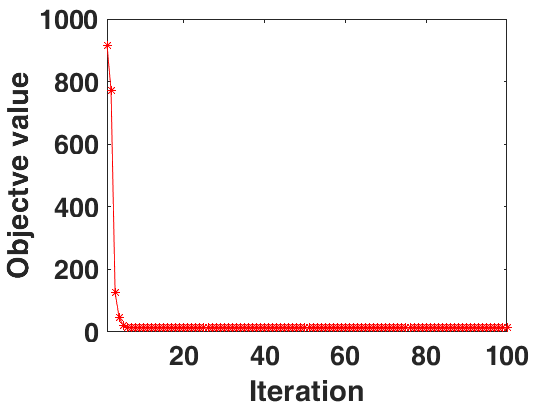}}
\hspace{0.0cm}
\subfigure[HDED]{\label{Conv_dreamer}\includegraphics[width=0.151\textwidth]{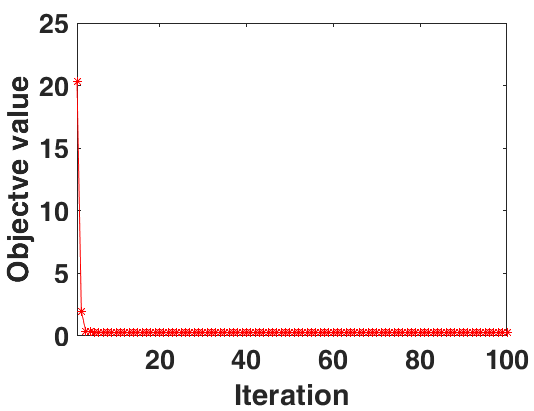}}
\caption{The convergence of the CWEFS algorithm.}
\label{conv_res}
\end{figure}

\subsection{Parameter sensitivity and convergence analysis}
CWEFS incorporates five tradeoff parameters in its objective function. We systematically evaluate parameter sensitivity through controlled experiments with the following configurations: the values of $\lambda$, $\beta$, $\eta$, $\gamma$, and $\delta$ are each adjusted within the set $\left\{ 0.001, 0.01, 0.1, 1, 10, 100, 1000 \right\}$, while $\gamma$ is varied within $\left\{ 2, 3, 4, 5, 6, 7, 8 \right\}$. With four parameters fixed at 0.1 (2 for $\gamma$), the remaining parameter is searched within its respective ranges. Owing to space constraints, only the sensitivity results for DEAP are presented. The 3D histogram in Fig.~\ref{Results_bar3_deap} quantifies AP stability. As illustrated in Fig.~\ref{Results_bar3_deap}, AP remains nearly unchanged as other parameters vary, indicating that the performance of CWEFS on DEAP is not highly sensitive to the tradeoff parameters.

Furthermore, we validated the convergence properties of the CWEFS algorithm. Fig.~\ref{conv_res} shows convergence curves of the objective value with $\gamma=2$ and other parameters fixed at 10. As depicted in Fig.~\ref{conv_res}, CWEFS converges rapidly within a small number of iterations, demonstrating the effectiveness of the optimization strategy.

\section{Conclusions}\label{Conclusion}
Inspired by brain volume conduction effects, a channel-wise EEG feature selection model is proposed for the multi-dimensional emotion recognition, integrating shared latent structure learning and adaptive channel-weight learning to address two critical challenges: (1) quantifying channel-specific contributions to emotion recognition through dynamic weighting, and (2) discovering a unified latent representation that bridges multi-channel EEG spaces and multi-dimensional emotional space. CWEFS is optimized using an iterative algorithm that ensures convergence and computational efficiency. Experimental results across three datasets demonstrate that CWEFS effectively selects informative EEG features for the emotion recognition task.

\bibliography{aaai2026}

\end{document}